\documentclass[10pt,polish,american]{article}
\usepackage[T1]{fontenc}
\usepackage[utf8]{inputenc}
\usepackage[a4paper]{geometry}
\usepackage{color}
\usepackage{wrapfig}
\usepackage{amstext}
\usepackage{graphicx}
\usepackage{setspace}
\usepackage{natbib}
\PassOptionsToPackage{normalem}{ulem}
\usepackage{ulem}
\definecolor{LightGray}{gray}{0.9}
\definecolor{LightGray1}{gray}{0.8}
\definecolor{lightseagreen}{RGB}{32,178,170}
\definecolor{gold3}{RGB}{205,173,0}
\definecolor{gold4}{RGB}{139,117,0}

\def\degr{\hbox{$^\circ$}}

\def\farcs{\hbox{$.\!\!^{\prime\prime}$}}

\usepackage{babel}
\begin{document}

\title{{\bf{ Comet C/2013 A1 Siding Spring. \\How treatment of data and NG\,effects can
change our predictions about close encounters with Mars ?}}}

\author{{{Pawe\l  ~Wajer}\thanks{E-mail: wajer@cbk.waw.pl} ~and {Ma\l gorzata Kr\'olikowska$^1$}\thanks{E-mail:
mkr@cbk.waw.pl}} \\
\footnotesize{Space Research Centre of the Polish Academy of Sciences, Bartycka 18A, 00-716 Warsaw, Poland} \\
}

\maketitle

%\date{2014 October 7}
%\label{firstpage}

\begin{abstract}
{We show that the estimates of close encounter of this comet with Mars depend on data treatment.
Using the data taken in the two-year period, we derived that the comet will miss Mars on 2014 October~19 at the distance of about $140\,150\pm 55$\,km or $140\,300\pm 45$\,km from its center, depending on the method of data processing in the purely gravitational model of motion (based on non-weighted data or weighted data, respectively). Unfortunately, the non-gravitational model of motion is still very uncertain, thus we can only speculate about estimates of expected distances for non-gravitational orbital solutions. However, we did not obtain a significant differences in close encounter prediction between the non-gravitational solutions and the gravitational ones.} \end{abstract}

%\begin{keywords}
%keywords: Solar system :general, Oort Cloud, comets:C/2013 A1 Siding Spring
%\end{keywords}
\vspace{0.3cm}

\section{\label{sec:introduction}Introduction}

This is the update of our study of the near-future dynamical evolution of comet C/2013 A1 Siding Spring. Comet Siding Spring comes from the Oort spike (with $1/a_{\rm ori}$ about $34-39\> \times 10^{-6}$\,au$^{-1}$ for purely gravitational model of motion, see Table~\ref{tab:models}) and experiences close encounters with Mars on 2014~October~19. Our preliminary solutions were based on the 713~positional measurements from 2012 October~4 to 2014 March~3. At that time, the comet was more than 3\,AU from the Sun and non-gravitational (NG) effects were indeterminable in its motion. Thus, to include NG effects to our analysis, we constructed a grid of various radial (described by constant parameter $A_1$) and/or transverse ($A_2$) components of NG acceleration using standard NG~model of motion given by Marsden, Sekanina and Yeomans (1973, hereafter MSY). These preliminary results were presented at the beginning of July in Helsinki at ACM conference \citep{wajer-krolikowska:2014} and also are reproduced in Fig 1.

Here, we show the results based on six months longer sequence of astrometric data (up to September 21). We analyze how selection and weighting of observations influence on the estimates of the comet position during its close encounter with Mars. We also determine very preliminary NG~orbit using MSY model -- it is important to stress however, that the comet is still on ingoing leg of its orbit, and NG~solutions are very uncertain. Additionally, to better understand future close approach to Mars, for each of models described below, a sample of 5\,001 VOs (including the nominal orbit) was constructed by using method given by Sitarski (1998).

\section{\label{sec:GR_model}Purely gravitational osculating orbits}

We present here three purely gravitational models, which differ in the data set treatment. Solution~G1 (see Table~\ref{tab:elements}) is derived using only selection procedure, solution~G2 is based on data selected and weighted in an iterative process of orbit determination (for more details see \citet{krolikowska-sit-soltan:2009}), and solution~G3 is obtained using the normal places instead of some series of measurements taken on the same day; therefore, in this last GR~model the number of residuals taken for orbit determination was drastically reduced from more than 2\,000 to 577. Table~\ref{tab:elements} shows all these three osculating orbital solutions, the numbers of residuals taken for orbit determination are also given there.

{\small{
\begin{center}
\begin{table}
\caption{\label{tab:elements} Gravitational heliocentric osculating orbits  based on the data interval of 2012 October~4--2014 March~3 (1159 measurements) reported to the \citet{IAU_MPC_Web} as of 2014 September 27. Angular elements are referred to Equinox J2000. Numbers in parenthesis indicate the 1$\sigma$ formal uncertainties of the last few corresponding digits.}
\begin{tabular}{lccc}
  \hline
                & \textcolor{red}{\bf Solution G1} & \textcolor{blue}{\bf Solution G2} & \textcolor{green}{\bf Solution G3}   \\
Data treatment  & \textcolor{red}{\bf non-weighted}& \textcolor{blue}{\bf weighted}    & \textcolor{green}{\bf normal places} \\
\hline \\
Epoch                           & 2014 Oct. 1.0             & 2014 Oct. 1.0        &     2014 Oct. 1.0      \\
Time of perihelion passage      & 20141025.314735 (43)      & 20141025.314920 (31) & 20141025.314665 (84)   \\
Perihelion distance (au)        & 1.39870303 (45)           & 1.39870150 (37)      & 1.39870659 (107)       \\
eccentricity                    & 1.00088460 (83)           & 1.00088020 (56)      & 1.00088676 (155)       \\
Argument of perihelion (\degr ) & 2.434593 (37)             & 2.434774 (27)        & 2.434396 (75)          \\
Longitude of node (\degr )      & 300.977178 (11)           & 300.977221 (8)       & 300.977104 (26)        \\
Inclination  (\degr )           & 129.027406 (5)            & 129.027397 (4)       & 129.027394 (11)        \\
& & & \\
RMS                             & 0\farcs 56                & 0\farcs 42           & 0\farcs 56          \\
Number of residuals             & 2250                      & 2264                 & 577           \\
\hline
\end{tabular}
\end{table}
\end{center}
}}

\section{\label{sec:NG_model}Numerical approach to nongravitational perturbations}

To determine the NG~cometary orbit the standard formalism described by MSY was used, where NG~acceleration can be written as

\begin{eqnarray}
F_{i}=A_{\rm i}\cdot \underbrace{\alpha\left(r/r_{0}\right)^{-2.15}\left[1+\left(r/r_{0}\right)^{5.093}\right]^{-4.614}}_{g(r){\rm - \>function\>symmetric\>relative\>to\>perihelion}},\qquad A_{\rm i}={\rm ~const~~for}\quad{\rm i}=1,2,3,
\end{eqnarray}

\noindent where $F_{1},\, F_{2},\, F_{3}$ are the radial, transverse and normal components
of the NG~acceleration, respectively, and the radial acceleration is defined outward along the Sun-comet line.
The normalization constant $\alpha=0.1113$ gives $g(1$~AU$)=1$; the scale distance $r_{0}=2.808$~AU.

An asymmetric model of NG~acceleration can be derived by using $g(r(t-\tau))$ instead of $g(r(t))$
\citep{yeomans-chodas:1989,sitarski:1994b}. This last model introduces an additional NG~parameter $\tau$ -- the time displacement of the maximum of the $g(r)$ relative to the moment of perihelion passage.
We found that only the symmetric model of NG~motion can be applicable for C/2013 A1 at this moment (see below).

\vspace{0.1cm}

Here, two NG~solutions are presented for two different data treatment: NG~model based on non-weighting data,
and NG~model derived using weighting procedure for the data set (Model A); in both cases three NG~parameters were determined, where
parameter $A_1$ is negative (solutions with assumed $A_3=0$ also give negative $A_1$). We noticed, however, no improvement in the fit to the data in comparison to the purely gravitational solutions.
We found that asymmetric NG~solutions with the maximum of the $g(r)$ shifted more than 200 days before perihelion gives positive $A_1$ and slightly smaller RMS, however such a solution seems to be even more exotic than two symmetric models presented here.
We also obtained the NG~model using normal places instead of some original measurements. This solution, however, gives even more negative radial component of NG~acceleration. Therefore, the results based on this model are not presented here.
We also determined the NG~solution (on the basis of 1159~measurements taken up to September~21) using the parameters $A_1$ and $A_2$ given by Farnocchia at \cite{JPL_Browser} (available on October~1) and based on 180~observations selected by him from more than 2\,000~taken in the period 2012~Oct.~4--2014~Sept.~26 (Model B).

\vspace{0.1cm}

To determined the more realistic NG~solution for  C/2013~A1 we probably have to wait a few months. For example, at \cite{IAU_MPC_Web} only purely gravitational orbital solution was presented at this moment (October~1).

%\vspace{0.2cm}

%The equations of a comet's motion were integrated numerically using the recurrent power series method, taking into account perturbations by all the planets (additionally, Pluto was taken into account to be consistent with DE405/WAW) and including the relativistic effects. All orbital calculations performed for this catalogue were based on the Warsaw numerical ephemeris DE405/WAW of the Solar System (Sitarski 2002), consistent with a high accuracy with the JPL ephemeris DE405.

\section{\label{sec:introduction} Close encounter with Mars}

Results of our investigations are summarized in the Table 2 and Figure 2, where three solutions are based on purely gravitational motion and differ in data treatment (as was described above) and the next set of three solutions include NG~acceleration in the model of motion.

\begin{center}
\begin{table}[htbp]
{\footnotesize{
\caption{\label{tab:models} Gravitational (GR) and non-gravitational (NG) solutions based on the data interval of 2012 October~4--2014 March~3.In the \textcolor{magenta}{\bf third NG~model}, parameters $A_1$ and $A_2$ was assumed according to Farnocchia solution given at \citet{JPL_Browser} at the end of September; this NG~determination was based on 180 measurements selected by Farnocchia from more than 2000 observations. }
\begin{tabular}{lcccccccc}
  \hline
 Data       & $A_1$ &  $A_2$   &  $A_3$   & Nominal distance & $\mu$ & $\sigma$ & $1/a_{\rm ori}$ & $1/a_{\rm fut}$  \\
 treatment  & \multicolumn{3}{c}{in units of $10^{-8}$AU/day$^2$} & \multicolumn{3}{c}{i n ~~u n i t s ~~of~~km} & \multicolumn{2}{c}{in units of $10^{-6}$AU$^{-1}$} \\
\hline \\
\multicolumn{9}{c}{{\bf Gravitational orbital solutions }} \\
\textcolor{red}{\bf non-weighted}   & --         &   --      &   --      & 140150  & 140148  &51 & 35.66$\pm$0.60 &  115.6$\pm$0.6 \\
\textcolor{blue}{\bf    weighted}   & --         &   --      &   --      & 140299 &140300  & 42 & 38.58$\pm$0.41 &  118.1$\pm$0.4 \\
\textcolor{green}{\bf normal places}& --         &   --      &   --      & 139840 &139843  & 125 & 34.38$\pm$1.08 &  115.2$\pm$1.0 \\
\\
\multicolumn{9}{c}{{\bf Non-gravitational orbital solutions}} \\
\textcolor{gold3}{\bf non-weighted}  & $-1.69\pm 0.36$ & $1.22\pm 0.66$ & $-0.33\pm 0.37$ & 141099 & 141125  &728 & 37.80$\pm$1.12 &  8.8 $\pm$51.3 \\
\textcolor{cyan}{\bf    weighted}  & $-0.92\pm 0.30$ & $0.33\pm 0.53$ & $-0.49\pm 0.29$ &141520  & 141569  & 586 & 38.90$\pm$0.80 & 91.0 $\pm$41.6 \\
\textcolor{magenta}{\bf weighted}  & 1.46 &   0.147    &   --      & 139603 & 139603 & 42 & 40.65$\pm$0.41  &    110.8$\pm$0.4 \\
  \hline
\end{tabular}
}}
\end{table}
\end{center}

\begin{figure}[htbp]
\includegraphics[width=13cm]{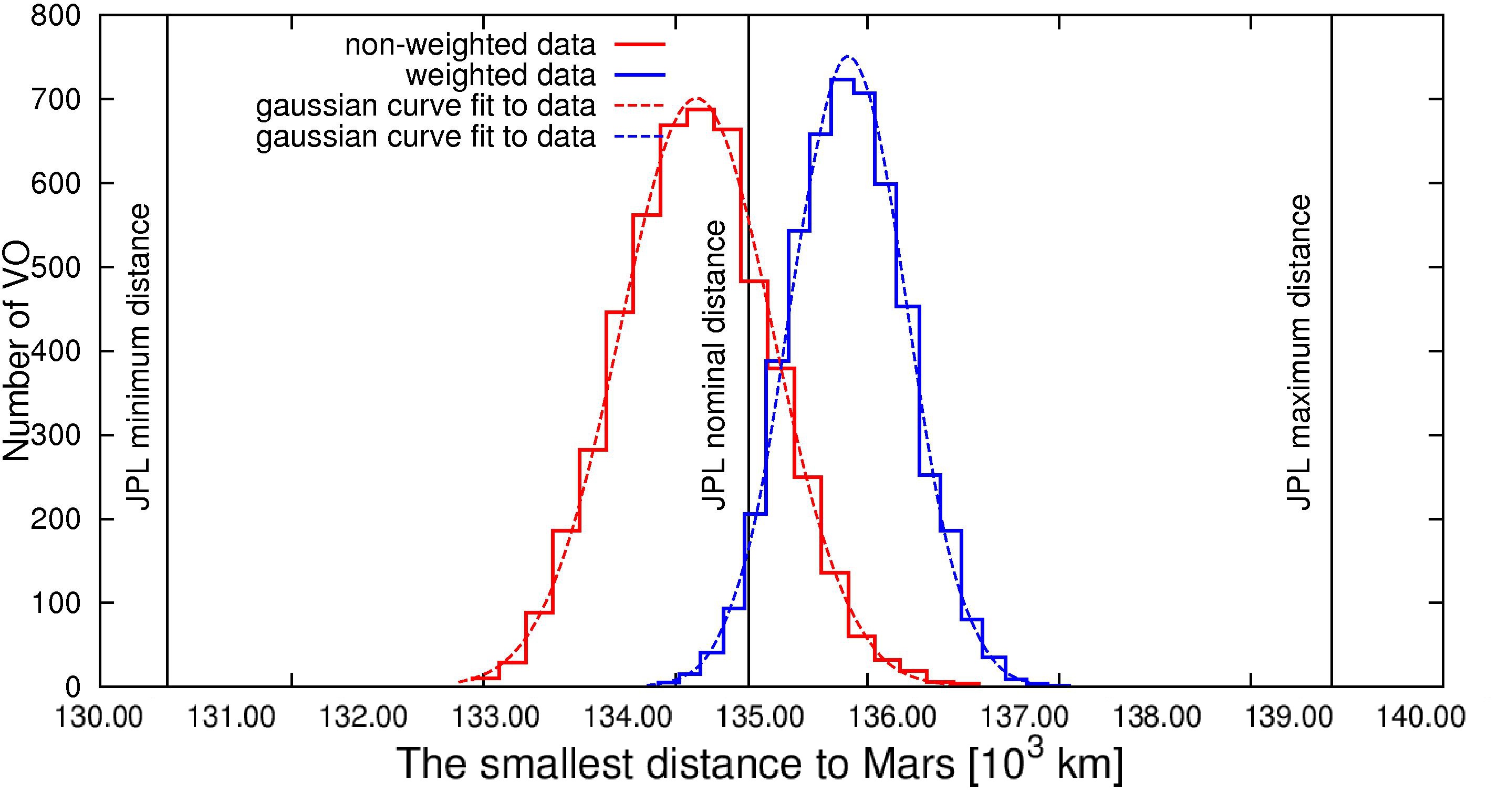}
\caption{\label{fig:GRcomparison}
\footnotesize{Cumulative distribution of all VCs and the nominal orbit of the comet C/2013\,A1 during the close approach to Mars for purely gravitational motion as predicted in March 2014 \citep{wajer-krolikowska:2014}. The calculations were based on the observational arc from the period 2012 October~4--2014 March~3, also JPL~limits (vertical lines) were based on the same data time interval.
\newline \textcolor{red}{\bf Red curve shows purely gravitational results for non-weighted data,}
\newline \textcolor{blue}{\bf blue curve -- also GR solution but using weighted data.}
\newline Gaussian fit to our numerical results are shown by dashed red and blue lines. }}
\end{figure}

\begin{figure}[htbp]
\includegraphics[width=13cm]{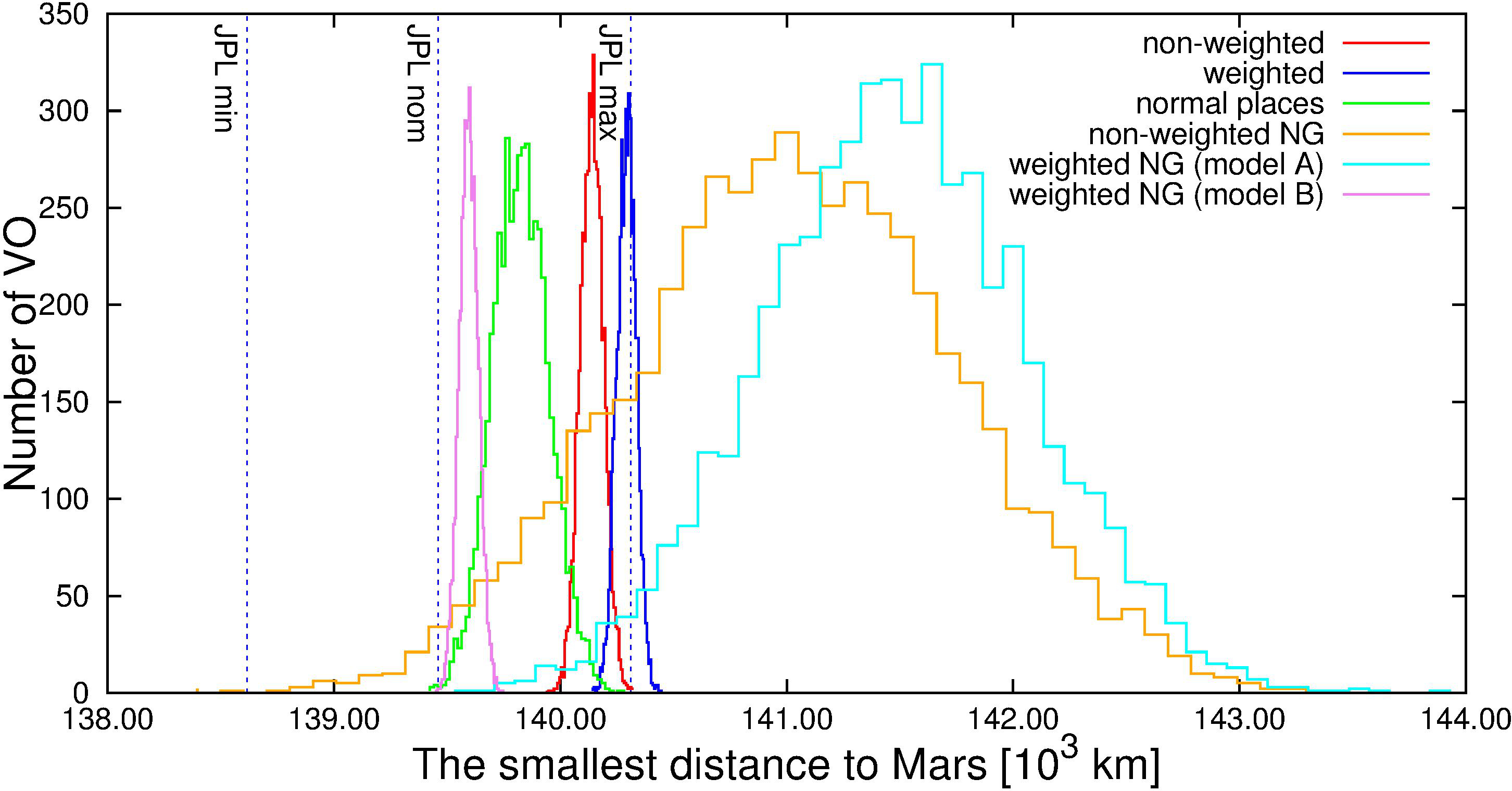}
\caption{\label{fig:NGcomparison}
\footnotesize{Cumulative distribution of all VCs and the nominal orbit of the comet C/2013\,A1 during the close approach to Mars for pure gravitational and for preliminary NG~models presented here. The calculations are based on the observational arc from the period 2012 October~4--2014 October 21, i.e. six months longer observational arc than in Fig.~\ref{fig:GRcomparison}. The vertical dotted straight lines show minimum, nominal and maximum distance of the comet from the Mars taken from JPL Small-Body Database on October~6.
\newline \textcolor{red}{\bf Red curve shows GR~results for non-weighted data (solution G1 in Table~\ref{tab:elements})},
\newline \textcolor{blue}{\bf blue curve -- GR solution using weighted data (solution G2 in Table~\ref{tab:elements})},
\newline \textcolor{green}{\bf green curve -- GR solution using normal places (solution G1 in Table~\ref{tab:elements})},
\newline \textcolor{gold3}{\bf gold curve -- NG solution using non-weighted data},
\newline \textcolor{cyan}{\bf cyan curve -- NG solution using weighted data},
\newline \textcolor{magenta}{\bf magenta curve -- NG solution using also non-weighted data, where parameters $A_1$ and $A_2$ were assumed and taken
from \citet[at the beginning of October]{JPL_Browser}.}}}
\end{figure}

\section{\label{sec:conclusions}Conclusions}
We analyzed the close approach of the comet C/2013 A1 (Siding Spring) with Mars by using the observational arc from 2012 October 4 to 2014 October 21. For detailed analysis of our models a sample of 5001 VO (including the nominal orbit)  was generated. We created three purely gravitational models based on non-weighted data, weighted data and using the normal places instead of some series of measurements taken on the same day. Though the non-gravitational model of motion is still very uncertain, we also analyzed how non-gravitational perturbations can change the close encounter predictions.
Presented calculations confirm the widespread opinion that the comet C/2013~A1 Siding Spring will not collide with Mars.
Our more detailed conclusions can be summarized as below:
\begin{enumerate}
\item
Using the purely gravitational models, we obtained that the nominal closest distance to Mars of the comet C/2013 A1 will be: 140\,150\,km\,$\pm$\,155\,km (non-weighted model), 140\,300\,$\pm$\,130\,km (weighted model) and 139\,840\,$\pm $\,380\,km in case of the solution based on the normal places, respectively ($3\sigma$ error estimates). We conclude that the comet may pass the Mars a little farther in comparison with \citet{Farnocchiaetal2014}  and our previous \citep[see also Fig.~\ref{fig:GRcomparison}]{wajer-krolikowska:2014} estimates.  
\item
Our GR~solutions as well as NG~solution with assumed NG~parameters are generally in a very good agreement with newer NG~solution present by Farnocchia at \citet{JPL_Browser}, though our estimate of formal uncertainties ($1\sigma$-error) seems smaller for each individual solution. Fig.~2 shows that the data treatment causes larger uncertainties of the close encounter with Mars than the uncertainties of each of individual GR~solution. 
\item
In the case of non-gravitational perturbations we did not obtain significant differences between gravitational solutions, as also was predicted by \citet{Farnocchiaetal2014}. Including NG~perturbations to our analysis, we noticed significantly larger values of uncertainties of close encounters
with Mars.
\end{enumerate}

{{\bf  Acknowledgements.} The osculating orbits were calculated using the numerical orbital package developed by Grzegorz Sitarski and the Solar System Dynamics and Planetology Group at SRC PAS. }

\vspace{0.2cm}

\bibliographystyle{aa}

\end{document}